\documentclass{article}
\usepackage[psamsfonts]{amssymb}
\usepackage{amsmath}
\usepackage{cite}
\author{M. Khorrami$^1$\footnote{mamwad@mailaps.org}\ \  and
M. Alimohammadi$^{2}$\footnote{alimohmd@ut.ac.ir}
\\ $^1$ {\small Department of Physics, Alzahra University, Tehran
19938-91167, Iran.}
\\ $^2$ {\small School of Physics,
University of Tehran,}
\\ {\small North Karegar Ave., Tehran, Iran.} }
\title{Large-$N$ behavior of the Wilson loops of generalized two-dimensional Yang-Mills theories }
\date{}
\begin{document}
\maketitle
\begin{abstract}
\noindent The large-$N$ limit of the expectation values of the
Wilson loops in the fundamental representation, corresponding to
two-dimensional U$(N)$ Yang-Mills and generalized Yang-Mills
theories on a sphere are studied. The behavior of the expectation
values of the Wilson loops both near the critical area and for
large areas are investigated. It is shown that the expectation
values of the Wilson loops at large areas behave exponentially
with respect to the area of the smaller region the boundary of
which is the loop; and for the so called typical theories, the
expectation values of the Wilson loops exhibit a discontinuity in
their second derivative (with respect to the area) at the critical
area.
\end{abstract}
\section{Introduction}
Two-dimensional Yang-Mills theory (YM$_2$) and generalized
Yang-Mills theories (gYM$_2$'s) have been a subject of extensive
study during recent years
\cite{1,2,3,4,5,6,7,a,b,c,10,9,19,12,27,d,20}. These are important
integrable models which can shed light on some basic features of
QCD$_4$. Also, there exists an equivalence between YM$_2$ and the
string theory. It was shown that the coefficients of the $1/N$
expansion of the partition function of SU$(N)$ YM$_2$ are
determined by a sum over maps from a two-dimensional surface onto
the two-dimensional target space.

The partition function and the expectation values of the Wilson
loops of YM$_2$ have been calculated in lattice- \cite{1,8} and
continuum-formulations \cite{4,9,19}. The partition function and
the expectation values of the Wilson loops of gYM$_2$'s have been
calculated in \cite{c,10}. All of these quantities are described
as summations over the irreducible representations of the
corresponding gauge group. In general, it is not possible to
perform these summations explicitly. For large gauge groups,
however, these summations may be dominated by some specific
representations, and it can be possible to perform the summations
explicitly. There are other physical reasons as well (for example
the relation between large-$N$ YM$_2$ and the string theory), that
make the study of the YM$_2$ and gYM$_2$'s for large groups
important.

In \cite{11}, the large-$N$ limit of the U$(N)$ YM$_2$ on a sphere
was studied. There it was shown that the above mentioned summation
is replaced by a (functional) integration over the continuous
parameters of the Young tableaux corresponding to the
representation. Then the saddle-point approximation singles out a
so-called classical representation, which dominates the
integration. In this way, it was shown that the free energy of the
U$(N)$ YM$_2$ on a sphere with the surface area
$A<A_{\mathrm{c}}=\pi^2$ has a logarithmic behavior \cite{11}. In
\cite{12}, the free energy was calculated for areas $A>\pi^2$,
from which it was shown that the YM$_2$ on a sphere has a
third-order phase transition at the critical area
$A_{\mathrm{c}}=\pi^2$, like the well known Gross-Witten-Wadia
phase transition for the lattice two dimensional multicolour gauge
theory \cite{13,14}. The phase structure of the large-$N$ YM$_2$,
generalized YM$_2$'s, and nonlocal YM$_2$ on a sphere were further
discussed in \cite{27,20,37,39}. It is also seen that for surfaces
with no boundaries, only the genus-zero surface (i.e. the sphere)
has a nontrivial saddle-point approximation, and all other
surfaces have trivial large-$N$ behavior \cite{11}.

In \cite{e,f,g}, the expectation value of the Wilson loops
corresponding to the large-$N$ U($N$)-YM$_2$ on a sphere was
investigated. The study was again based on the fact that in the
large-$N$ limit, a certain representation of the group is singled
out in the partition function and the expectation value of any
quantity is essentially the value of that quantity calculated at
that representation. Using this, the behavior of the expectation
values of the Wilson loops for large-$N$ YM$_2$ was investigated
for small and large areas.

In this paper we want to study the expectation-value of the Wilson
loops in the fundamental representation, corresponding to
large-$N$ YM$_2$ and gYM$_2$'s on a sphere. In section 2, the
expectation values of a Wilson loop is obtained in terms of the
the area enclosed by the loop and the density of the dominant
representation. In section 3, the behaviors of the free energy and
the expectation values of the Wilson loops for areas near the
critical value are investigated by some general arguments. In
section 4, we justify explicitly the results of section 3 for two
special cases YM$_2$ and $G(z)=z^4$ gYM$_2$ models. Finally, in
section 5 the large-area behavior of the expectation value of the
Wilson loops is studied.
\section{The expectation values of the Wilson loops}
Following \cite{10,20}, a generalized U$(N)$ Yang-Mills theory on
a surface is characterized by a function $\Lambda$:
\begin{equation}\label{w1}
\Lambda(R)=\sum_{k=1}^p\frac{a_k}{N^{k-1}}\,C_k(R),
\end{equation}
where $R$ denotes a representation of the group U$(N)$, $a_k$'s
are constants, and $C_k$'s are the Casimirs of the group defined
through
\begin{equation}\label{w2}
C_k=\sum_{i=1}^N[(n_i+N-i)^k-(N-i)^k].
\end{equation}
$n_i$'s are nonincreasing integers characterizing the
representation. It is assumed that $p$ is even and $a_p>0$. For
simplicity, from now on it is further assumed that all $a_k$'s
with odd $k$'s vanish. The partition function of such a theory on
an orientable genus $g$ surface with $n$ boundaries is
\begin{equation}\label{w3}
Z_{g,n}(A;U_1,\dots,U_n)=\sum_R d_R^{2-2g-n}\,\chi_R(U_1)\cdots
\chi_R(U_n)\,\exp[-A\,\Lambda(R)],
\end{equation}
where $U_i$'s are the holonomies on the boundaries, $A$ is the
area of the surface, $d_R$ is the dimension of the representation
$R$, and $\chi_R$ is the character of the representation $R$.

Consider a simple (not self-crossing) loop, which divides a sphere
into two regions of areas $A_1$ and $A_2$. The expectation value
of the Wilson loop corresponding to this loop is (following
\cite{10,20,e})
\begin{equation}\label{w4}
W_{r}(A_1,A_2)=\frac{1}{Z_0(A_1+A_2)}\,\int\mathrm{d} U\;
Z_{0,1}(A_1,U)\,Z_{0,1}(A_2,U^{-1})\,\frac{1}{d_r}\,\chi_r(U),
\end{equation}
where $r$ is a representation of U$(N)$, $Z_0(A)$ is the partition
function on the sphere, and the integration is done over the
group. Eq. (\ref{w4}) can be rewritten as
\begin{equation}\label{w5}
W_{r}(A_1,A_2)=\frac{1}{d_r\,Z_0(A_1+A_2)}\sum_{R,S}d_R\, d_S\,
\langle R,r|S\rangle\,\exp[-A_1\,\Lambda(R)-A_2\,\Lambda(S)],
\end{equation}
where the summations run over irreducible representations of
U$(N)$, and $\langle R,r|S\rangle$ is the number of
representations $S$ in the tensor product of the representations
$R$ and $r$. From now on, the representation $r$ is taken to be
the fundamental representation, the dimension of which is $N$.

For large $N$, the summation on $R$ is dominated by the so called
classical representation, which maximizes the product
$d_R\,\exp[-A_1\,\Lambda(R)]$, as was shown in, for example,
\cite{20}. To obtain this representation, it is convenient to
introduce the new parameters
\begin{align}\label{w6}
x&:=\frac{i}{N},\nonumber\\
n(x)&:=\frac{n_i}{N},\nonumber\\
h(x)&:=-n(x)-1+x,
\end{align}
the density
\begin{equation}\label{w7}
\rho(h):=\frac{\mathrm{d} x}{\mathrm{d} h},
\end{equation}
and the function
\begin{equation}\label{w8}
G(z):=\sum_{k=0}^p a_k\,(-z)^k.
\end{equation}
Then the density corresponding to the classical representation is
characterized by
\begin{align}\label{w9}
\frac{A}{2}\,G(z)-\int\mathrm{d}
y\;\rho(y)\;\ln|z-y|&=\hbox{const.},\qquad \mathrm{iff}\;
\rho(z)\ne 1\hbox{
and } -a\leq z\leq a,\nonumber\\
\int_{-a}^a\mathrm{d} z\;\rho(z)&=1,
\end{align}
where $a$ is a positive number to be determined through the above
conditions.

Then, using arguments similar to those used in \cite{e}, one
arrives at
\begin{equation}\label{w10}
W(A_1,A_2)=\int_{-a}^a\frac{\mathrm{d}
z}{\pi}\;\sin[\pi\,\rho(z)]\;\exp\left[-\mathrm{P}\int_{-a}^a
\mathrm{d} y\;\frac{\rho(y)}{z-y}+A_2\,G'(z)\right],
\end{equation}
where $\mathrm{P}$ means Cauchy's principal value, or
\begin{equation}\label{w11}
W(A_1,A_2)=-\oint_C\frac{\mathrm{d} z}{2\pi
i}\;\exp\left[-\int_{-a}^a \mathrm{d}
y\;\frac{\rho(y)}{z-y}+A_2\,G'(z)\right].
\end{equation}
$G'(z)$ denotes the derivative of $G(z)$, and $C$ is a
counterclockwise contour encircling the real segment $[-a,a]$.
Defining the function $H$ of a complex variable $z$ as
\begin{equation}\label{w12}
H(z):=\int_{-a}^a\mathrm{d}y\;\frac{\rho(y)}{z-y},
\end{equation}
(as it was defined in \cite{20}), one can rewrite (\ref{w11}) as
\begin{equation}\label{w13}
W(A_1,A_2)=-\oint_C\frac{\mathrm{d}z}{2\pi
i}\;\exp\left[-H(z)+A_2\,G'(z)\right],
\end{equation}
It was seen in \cite{20} that $H$ is analytic except on the real
segment $[-a,a]$. In Yang-Mills case where $G(z)=(1/2)z^2$,
(\ref{w13}) leads to one derived in \cite{e,f}.
\section{Free energy and the expectation values of the Wilson loops near the critical
area} As it was discussed in \cite{12,20}, if the area of the
sphere is less than the critical value $A_{\mathrm{c}}$, then the
density $\rho$ is always less than 1 and the free energy of the
system is a smooth function of the area. For areas larger than
$A_{\mathrm c}$, however, $\rho$ becomes equal to 1 on certain
segments. This produces a discontinuous behavior in the
area-dependence of the free energy, at $A=A_{\mathrm{c}}$. There
was an argument in \cite{39} to obtain the degree of this
nonsmoothness , i.e. the order of this phase transition. However,
the argument is problematic and, as will be shown here, leads to
incorrect results at some cases.

Let us denote the correct density for $A>A_{\mathrm{c}}$ (the
strong region) by $\rho_s$ and the density satisfying the
saddle-point equation, but not subject to not exceeding 1, by
$\rho_w$.  Define $L_w$ and $L_s$ as the regions where $\rho_s$ is
less than one or equal to one, respectively; and $L'_w$ and $L'_s$
as the regions where $\rho_w$ is less than one or greater than
one, respectively. The length of $L_s$ is denoted by $2b$, so the
length of $L'_s$ is of the order $2b$. We have
\begin{align}\label{w14}
\frac{A}{2}\,G'(z)&=\mathrm{P}\int\mathrm{d}y\;\frac{\,\rho_s(y)}{z-y}
,\quad z\in L_w,\\ \label{w15}
\frac{A}{2}\,G'(z)&=\mathrm{P}\int\mathrm{d}y\;\frac{\,\rho_w(y)}{z-y}
,\quad z\in L'_w\cup L'_s.
\end{align}
where $\rho_s$ ($\rho_w$) has been defined zero outside $L_w\cup
L_s$ ($L'_w\cup L'_s$). Defining
\begin{equation}\label{w16}
\alpha:=\max(\rho_w)-1,
\end{equation}
and using the ansatz
\begin{equation}\label{w17}
\rho_{s0}(y):=\min[\rho_w(y),1]
\end{equation}
as an approximate solution to the eq.(\ref{w14}), it is seen that
the difference of the left-hand side with the right-hand side
calculated at $\rho_s=\rho_{s0}$, is of the order $\alpha$ when
the distance of $z$ from $L_s$ is of the order $b$, and of the
order $\alpha\,b$ when the distance of $z$ from $L_s$ is large
compared to $b$. To see this, one notes that
\begin{align}\label{w18}
D(z):=&\frac{A}{2}\,G'(z)-\mathrm{P}\int\mathrm{d}y\;
\frac{\,\rho_{s0}(y)}{z-y},\nonumber\\
=&\mathrm{P}\int\mathrm{d}y\;\frac{\,\rho_{w}(y)-\rho_{s0}(y)}{z-y},\nonumber\\
=&\mathrm{P}\int_{L'_s}\mathrm{d}y\;\frac{\,\rho_{w}(y)-1}{z-y},\quad
z\in L'_w.
\end{align}
The numerator of the integrand is of the order $\alpha$, and the
length of the integration region is of the order $b$. If the
distance of $z$ from $L_s$ (or $L'_s$) is of the order $b$, then
the denominator of the integrand is of the order $b$, and the
integral would be of the order $\alpha$. If the distance of $z$
from $L_s$ (or $L'_s$) is large compared to $b$, then the
denominator of the integrand is of the order one, and the integral
would be of the order $\alpha\,b$.

Defining $\delta\rho_s$ through
\begin{equation}\label{w19}
\rho_s=:\rho_{s0}+\delta\rho_s,
\end{equation}
it is seen that
\begin{equation}\label{w20}
D(z)=\mathrm{P}\int\mathrm{d}y\;\frac{\,\delta\rho_{s}(y)}{z-y},
\quad z\in L_w.
\end{equation}
$\delta\rho_s$ is at most of the order $\alpha$. But $\delta\rho$
cannot be of order $\alpha$ everywhere. In fact, if it is of the
order $\alpha$ in a region large compared to $b$, then the
right-hand side would be of the order $\alpha$ for some points $z$
the distance of them from $L_s$ (or $L'_s$) is large compared to
$b$. But $D(z)$ is of the order $\alpha\,b$ when the distance of
$z$ from $L_s$ (or $L'_s$) is large. So,
\begin{equation}\label{w21}
\rho_s(z)-\rho_w(z)=\begin{cases}
                                 O(\alpha),& z\in L''_s\\
                                 o(\alpha),& z\not\in L''_s
                    \end{cases},
\end{equation}
where $L''_s$ is a region around $L_s$, the length of which is of
the order $b$.

Using (\ref{w21}) and
\begin{equation}\label{w22}
H_{s,w}(z):=\int\mathrm{d}y\;\frac{\rho_{s,w}(y)}{z-y},
\end{equation}
it is seen that
\begin{equation}\label{w23}
H_s(z)-H_w(z)=\int_{L_s}\mathrm{d}y\;\frac{1-\rho_w(y)}{z-y}
+\int_{\mathbb{R}\backslash L_s}\mathrm{d}y\;
\frac{\delta\rho_s(y)}{z-y}=o(\alpha),\quad |z|\gg a,
\end{equation}
where $a$ is of the order of the length of the region in which the
integrand does not vanish.

For large values of $z$, the difference $H_s(z)-H_w(z)$ is an
analytic even function of $b$ \cite{20}. This means that this
difference can be expanded like
\begin{equation}\label{w24}
H_s(z)-H_w(z)=c_k(z)\,b^{2k}+c_{k+1}(z)\,b^{2k+2}+\cdots,
\end{equation}
where $k$ is a positive integer. If at the point $\rho_w$ attains
its maximum, the $2m$'th derivative of $\rho_w$ is its first
nonvanishing derivative, then $b^{2m}$ is of the order $\alpha$.
As the difference $H_s(z)-H_w(z)$ is $o(\alpha)$, it turns out
that $k$ must be greater than $m$. So one arrives at
\begin{equation}\label{w25}
H_s(z)-H_w(z)\sim b^{2m+2},\quad \hbox{for large }z
\end{equation}
or
\begin{equation}\label{w26}
H_s(z)-H_w(z)\sim \alpha^{1+(1/m)},\quad \hbox{for large }z.
\end{equation}

The free energy $F$ is defined through
\begin{equation}\label{w27}
F:=-\frac{1}{N^2}\ln Z,
\end{equation}
and one has \cite{20}
\begin{equation}\label{w28}
F'(A)=\int_{-a}^a\mathrm{d}y\;\rho(y)\,G(y).
\end{equation}
Using (\ref{w12}), one can rewrite this as
\begin{equation}\label{w29}
F'(A)=\oint_{C_\infty}\frac{\mathrm{d}z}{2\pi i}\;H(z)\,G(z),
\end{equation}
where $C_\infty$ is a large counterclockwise circle. Now, using
(\ref{w26}), it is seen that
\begin{equation}\label{w30}
F'_s(A)-F'_w(A)\sim (A-A_{\mathrm{c}})^{(2l-1)[1+(1/m)]},
\end{equation}
and from that
\begin{equation}\label{w31}
F_s(A)-F_w(A)\sim (A-A_{\mathrm{c}})^{1+(2l-1)[1+(1/m)]},
\end{equation}
where the $(2l-1)$'th derivative of $\alpha$ with respect to $A$
is the first nonvanishing derivative of $\alpha$ with respect to
$A$, so $\alpha(A)\sim (A-A_{\mathrm{c}})^{2l-1}$. It is seen that
in general the last equation differs that found in \cite{39}. The
two, however, coincide if $l=1$. The flaw in the argument of
\cite{39} is that from the fact that the integral of the
difference $\rho_s-\rho_w$ on $L_w$ is of order $\alpha\,b$, it
was deduced that $\rho_s-\rho_w$ itself is of order $\alpha\,b$,
which is not true, as seen from (\ref{w21}). However, for typical
theories where $l=m=1$, it is still true that the system exhibits
a third order phase-transition at $A=A_{\mathrm{c}}$.

Using (\ref{w13}) and (\ref{w26}), one can also investigate the
nonsmoothness behavior of the expectation values of the Wilson
loops. Using (\ref{w13}), It is seen that
\begin{equation}\label{w32}
W_s-W_w\sim (A-A_{\mathrm{c}})^{(2l-1)[1+(1/m)]},
\end{equation}
so that for typical theories, there is a second order
discontinuity in the expectation values of the Wilson loops. For
example for all gYM$_2$ theories with $G(z)=a_k\,z^{2k}(k\geq 1)$,
including YM$_2$, it has been shown that \cite{39}
\begin{equation}\label{w33}
\alpha(A)=\left( \frac{A}{A_{\mathrm c}}\right)^{\frac{1}{2k}}-1.
\end{equation}
This shows that the derivative of $\alpha$ with respect to $A$
never vanishes, or $l=1$. Also the second derivative of $\rho_w$
is negative at the maximum of $\rho_w$, which gives $m=1$. So for
all theories with $G(z)=a_k\,z^{2k}$,
\begin{equation}\label{w34}
W_s-W_w\sim\left(A-A_{\mathrm c}\right)^2.
\end{equation}
\section{Explicit evaluation of the order of discontinuity of Wilson
loops for YM$_2$ and the gYM$_2$ with $G(z)=z^4$} From
(\ref{w12}), and using the fact that $\rho$ is even if $G$ is
even, one has
\begin{equation}\label{w35}
H(z)=\frac{1}{z}+\frac{1}{z^3}\int_{-a}^a\mathrm{d}y\;y^2\rho(y)
+\frac{1}{z^5}\int_{-a}^a\mathrm{d}y\;y^4\rho(y)+\cdots.
\end{equation}
Using this and (\ref{w13}), one arrives at
\begin{align}\label{w36}
W_s-W_w&=\oint_C\frac{\mathrm{d}z}{2\pi
i}\;\exp\left[-H_w(z)+A_2\,G'(z)\right]\,
[H_s(z)-H_w(z)+\cdots],\nonumber\\
&=\oint_C\frac{\mathrm{d}z}{2\pi
i}\;\exp\left[-H_w(z)+A_2\,G'(z)\right]\,\left[
\frac{\Delta(y^2)}{z^3}+\frac{\Delta(y^4)}{z^5}+\cdots\right],
\end{align}
where
\begin{align}\label{w37}
\Delta(y^n):=&\langle y^n\rangle_s-\langle
y^n\rangle_w,\nonumber\\
=&\int\mathrm{d}y\;y^n\rho(y)\,[\rho_s(y)-\rho_w(y)].
\end{align}
It is seen that $W_s-W_w$ is proportional to $\Delta(y^2)$. For
small values of $A_2$, one can expand $\exp[A\,G'(z)]$ and it is
seen that only a finite number of terms in (\ref{w36}) are needed
to obtain the behavior of $W_s-W_w$. (In the product of
$[G'(z)]^n\,[H_s(z)-H_w(z)]$, one needs to keep only terms of
$z^k$ for $k\geq -1$.)

For example, if $G(z)=(1/2)\,z^2$, using the first two terms one
can obtain the behavior of $W_s-W_w$ up to order $A_2^4$. For this
theory, one has (using \cite{12})
\begin{align}\label{w38}
\Delta(y^2)&=2(F'_s-F'_w),\nonumber\\
&=\frac{2}{\pi^2}\left( \frac{A-A_{\mathrm c}}{A_{\mathrm
c}}\right)^2+\cdots,
\end{align}
and
\begin{equation}\label{w39}
\Delta(y^4)=\frac{4}{\pi^2}\left(\frac{A-A_{\mathrm c}}{A_{\mathrm
c}}\right)^2+\cdots,
\end{equation}
which shows that leading term of $W_s-W_w$ is proportional to
$(A-A_{\mathrm{c}})^2$, at least up to order $(A_2)^4$.

As a second example, one can take $G(z)=z^4$. Following \cite{20},
one can expand $H_s-H_w$ near the critical area, and expand it in
terms of the powers of $z^{-1}$, from which $\Delta(y^2)$ (the
coefficient of $z^{-3}$) is calculated to be
\begin{equation}\label{w40}
\Delta(y^2)=\frac{551}{486\pi^2}\left(\frac{A-A_{\mathrm
c}}{A_{\mathrm c}}\right)^2+\cdots,
\end{equation}
which shows that the leading term of $W_s-W_w$ is proportional to
$(A-A_{\mathrm{c}})^2$, at least up to first order in $A_2$.

\section{The expectation values of Wilson loops for large areas}
To study the Wilson loop on the plane ($A\to\infty,$), one notes
that for large values of $A$, the first equation of the set
(\ref{w9}) cannot be fulfilled with finite $\rho$. This shows that
as $A$ tends to $\infty$, the density $\rho$ becomes equal to one
everywhere in $(-a,a)$. Then the second equation shows that $a$
tends to $(1/2)$ as $A$ tends to $\infty$:
\begin{align}\label{w41}
\lim_{A\to\infty}\rho(z)&=1,\qquad -a<z<a,\nonumber\\
\lim_{A\to\infty}a&=\frac{1}{2}.
\end{align}
This shows that
\begin{equation}\label{w42}
\lim_{A\to\infty}H(z)=\ln\frac{z+\frac{1}{2}}{z-\frac{1}{2}}.
\end{equation}
For $A$ large but not infinite, one can use (\ref{w12}) to expand
$H(z)$ around the value of $H$ for infinite $A$. The result would
be
\begin{equation}\label{w43}
H(z)=\ln\frac{z+\frac{1}{2}}{z-\frac{1}{2}}+\sum_{i,n}\frac{\alpha_{i,n}(A)}{(z-z_i)^n},
\end{equation}
where $n$ runs over positive integers and $z_i$'s are points in
$[-\frac{1}{2},\frac{1}{2}]$ around them $\rho$ differs from one
(for large but not infinite areas). Among these points are
$-\frac{1}{2}$ and $\frac{1}{2}$. $\alpha_{i,n}$'s should of
course tend to zero as $A$ tends to infinity. Using the above
expansion in (\ref{w13}), one arrives at
\begin{equation}\label{w44}
W(A_1,A_2)=\exp\left[A_2\,G'\left(-\frac{1}{2}\right)\right] +
\sum_i\beta_i(A_2,A)\,\exp[A_2\,G'(z_i)],
\end{equation}
where $\beta_i$'s are sums of polynomials in $A_2$ times functions
of $A$. One can write the above equation in a form more
manifestly-symmetric with respect to $A_1$ and $A_2$:
\begin{align}\label{w45}
W(A_1,A_2)=&\exp\left[A_2\,G'\left(-\frac{1}{2}\right)\right] +
\exp\left[A_1\,G'\left(-\frac{1}{2}\right)\right]\nonumber\\
&+ \sum_i\tilde\beta_i(A_2,A)\,\exp[A_2\,G'(z_i)],
\end{align}
where
\begin{equation}\label{w46}
\tilde\beta_i:=\begin{cases}
                            \beta_i,&z_i\ne\frac{1}{2}\\
                            \beta_i-\exp\left[A\,G'\left(-\frac{1}{2}\right)\right],&
                            z_i=\frac{1}{2}
               \end{cases}
\end{equation}
and use has been made of the fact that $G'$ is an odd function.

For YM$_2$, one can perform a more explicit calculation. Here it
is known that for $A>A_{\mathrm{c}}$, the density is equal to one
for $-b<z<b$, where $b$ is some positive number less than $a$.
Also, it is seen that $a$ and $b$ both tend to $\frac{1}{2}$ as
$A$ tends to $\infty$. So there are only two $z_i$'s and one has
\begin{equation}\label{w47}
H(z)=\ln\frac{z+\frac{1}{2}}{z-\frac{1}{2}}+\alpha(A)
\left[\frac{1}{(z-\frac{1}{2})^2}-\frac{1}{(z+\frac{1}{2})^2}\right]
+O(\alpha^2),
\end{equation}
where
\begin{equation}\label{w48}
\alpha=\frac{1}{2}\left(\frac{1}{2}-b\right)^2+
\int_b^a\mathrm{d}y\;\left(y-\frac{1}{2}\right)\,\rho(y).
\end{equation}
(One can see that $H$ is an odd function and its residue at $z=0$
should be equal to one, so there are no first-order poles at
$z=\pm\frac{1}{2}$.) Putting this in (\ref{w13}), one arrives at
\begin{align}\label{w49}
W=&-\oint\frac{\mathrm{d}z}{2\pi\,i}\,\frac{z-\frac{1}{2}}{z+\frac{1}{2}}\,
\left\{1-
\alpha\,\left[\frac{1}{(z-\frac{1}{2})^2}-\frac{1}{(z+\frac{1}{2})^2}\right]+
O(\alpha^2)\right\}\,\exp(A_2\,z),\nonumber\\
=&\exp\left(-\frac{A_2}{2}\right)\,\left[1+\alpha\,
\left(-1-A_2+\frac{A_2^2}{2}\right)\right]+
\alpha\,\exp\left(\frac{A_2}{2}\right)+O(\alpha^2).
\end{align}

It is known that for $A_2=A$, $W_s$ should become unity. This
shows that up to the leading order,
\begin{equation}\label{w50}
\alpha=\exp\left(-\frac{A}{2}\right).
\end{equation}
So,
\begin{equation}\label{w51}
W=\exp\left(-\frac{A_2}{2}\right)\,\left[1+\exp\left(-\frac{A}{2}\right)\,
\left(-1-A_2+\frac{A_2^2}{2}\right)\right]+
\exp\left(-\frac{A_1}{2}\right)+\cdots.
\end{equation}
This is the same relation obtained in \cite{e}, using a different
method.

\end{document}